\documentclass[10pt]{myiopart}
\usepackage{color} 
\usepackage{tabularx} 
\usepackage{epsfig}
\usepackage{bm}
\usepackage{graphicx}
\usepackage{epsfig}

\begin{document}

\title{\large Viewing Vanilla Quantum Annealing Through Spin Glasses}

\author{
Helmut G.~Katzgraber
}

\address{Department of Physics and Astronomy, Texas A\&M University,
College Station, Texas 77843-4242, USA; 1QB Information Technologies
(1QBit), Vancouver, British Columbia, Canada V6B 4W4; Santa Fe
Institute, 1399 Hyde Park Road, Santa Fe, New Mexico 87501, USA}

\ead{firstname@lastname.org}

\begin{abstract}

Quantum annealing promises to solve complex combinatorial optimization
problems faster than current transistor-based computer technologies.
Although to date only one commercially-available quantum annealer is
procurable, one can already start to map out the application scope of
these novel optimization machines. These mid-scale programmable analog
special-purpose devices could, potentially, revolutionize optimization.
However, their disruptive application domain remains to be found.  While
the commercial analog quantum optimization machine by D-Wave Systems
Inc.~already exceeds $1000$ qubits, here it is argued that maybe smaller
devices with better quality qubits, higher connectivity, and more
tunability might be better suited to answer if quantum annealing will
ever truly outperform specialized silicon technology combined with
efficient heuristics for optimization and sampling applications.

\end{abstract}

\ioptwocol

\noindent ``{\em The more you know, the more you can create. There's no
end to imagination in the kitchen.}'' --- Julia Child

\medskip

\noindent Discussing the medium-term impact of current quantum computing
devices is a nontrivial task given the different possible
implementations.  In this brief article some {\em food for thought} on
this subject is presented. It is noted that the expressed opinions are
those of the author alone.

\section{Amuse bouche --- Setting the stage}
\label{sec:intro}

What can we do with approximately $1000$ qubits? That depends strongly
on the {\em kind} of qubits.  There are different approaches to using
quantum mechanics in computation and, as the reader will notice, the
term {\em quantum computing} \cite{nielsen:00} is used very carefully
within this context. While the ultimate goal is to one day build a
digital programmable (universal) quantum computer that fully exploits
all the benefits of quantum entanglement and parallelism using $1000$
qubits, to date, we are far from this goal. Multiple
institutions---either academic, industrial, or governmental---have
recently invested heavily in quantum technologies. Small-scale few-qubit
programmable devices, such as IBM's publicly accessible superconducting
transmon device \cite{comment:ibm,castalvecchi:17} or a fully-connected five-qubit ion
trap device, \cite{linke:17} have been successfully used for
computations. However, their scope remains limited due to short
coherence times and a small number of qubits. Large corporations, such
as Google, IBM, Intel, NTT, and Microsoft, as well as smaller startups
such as Rigetti Computing have set ambitious goals
\cite{comment:supremacy}, however, it remains to be seen how useful
their devices will be in the near term.

History has shown that large digital computing revolutions are often
preceded by analog developments. The most paradigmatic example being
semi-programmable analog vacuum tube computing machines heralding the
development of programmable digital transistor technologies. In the
quantum computing world, analog semi-programmable quantum optimization
machines, such as those manufactured by D-Wave Systems Inc., will likely
be remembered as the precursor of programmable digital quantum devices.
Like vacuum tubes, or any other analog computing platform, their
application scope is limited, because they are designed with a
particular purpose in mind --- in this case, the minimization of the
cost function (Hamiltonian) of a binary (Boolean) optimization problem
\cite{hartmann:01,juenger:01,hartmann:04}. This does not immediately
imply that quantum annealing machines are doomed to eventually disappear
completely. To date, analog silicon devices are used as special-purpose
machines in many applications and will likely experience a renaissance
as co-processors now that Moore's Law \cite{moore:65} is apparently
slowly coming to an end. Furthermore, the development of (analog)
quantum optimization co-processors might find applications across
different areas of computing, ranging from optimization to machine
learning; time will tell.

D-Wave Systems Inc.~\cite{johnson:11} has pioneered the use of
superconducting flux qubits to build semi-programmable analog
optimization machines using transverse-field quantum annealing
\cite{kadowaki:98,finnila:94,farhi:01,santoro:02,das:05,santoro:06,das:08}.
Quantum annealing, the quantum counterpart of thermal simulated
annealing \cite{kirkpatrick:83} is a sequential optimization technique
where quantum fluctuations induced by a transverse field
\cite{das:05,das:08} are slowly quenched following an annealing
protocol, in the hope of minimizing the cost function of a quadratic
binary optimization problem. Although the performance of the device
remains controversial to date (see, for example,
Refs.~\cite{johnson:11,dickson:13,boixo:13a,katzgraber:14,boixo:14,ronnow:14a,katzgraber:15,heim:15,hen:15a,rieffel:15,mandra:16b,boixo:16,denchev:16,king:17}),
having direct access to the device has advanced quantum
computing substantially, as well as revolutionized how we think of
optimization today.

In this overview the focus will be placed on {\em quantum optimization}
and {\em quantum sampling} of cost functions of binary optimization
problems. While this scope might seem narrow at first, it is likely that
quantum annealing machines will play a central role in the field of
quantum optimization (and, more indirectly, in quantum computing) for
the next decade. Furthermore, optimization is ubiquitous in scientific,
as well as industrial applications. A plethora of optimization problems
can be mapped directly onto quadratic binary optimization problems
\cite{lucas:14}---which are the type of problem current quantum
annealing machines are designed to tackle---and so despite its seemingly
narrow scope, quantum optimization has the potential to revolutionize
machine learning, drug discovery, and industrial distribution, to name a
few.

The title of this overview cheekily provides a ``view of vanilla quantum
annealing through spin glasses.'' First, {\em vanilla} quantum annealing
refers to the implementation using a transverse field driver Hamiltonian
to minimize $2$-local cost functions, as it is done in the latest
version of the D-Wave Systems Inc.~D-Wave 2000Q quantum annealer. More
complex drivers could be used with potentially much better performance.
Among these could be non-stoquastic driver Hamiltonians with $2$-local
or even $4$-local symmetry. Similarly, multi-qubit native interactions
in the problem Hamiltonian beyond quadratic ($2$-local) would be
desirable for many applications.  However, these devices remain to be
built.  Second, spin glasses
\cite{mezard:87,young:98,binder:86,nishimori:01,stein:13} are likely the
hardest simple \cite{comment:simple} binary optimization problem. As
such, they are perfectly suited for benchmarking any new computing
paradigm aimed at minimizing a binary quadratic cost function. Not only
are they well understood, but a wide class of optimization problems can
be mapped directly onto spin-glass-like Hamiltonian \cite{lucas:14}.
Multiple recent studies that attempt to gain a deeper understanding of,
e.g., the D-Wave device have used variations of spin-glass Hamiltonians.
Carefully-designed Ising spin-glass problems can be used to probe the
existence of any quantum advantage (see, for example,
Refs.~\cite{lanting:14,katzgraber:15,denchev:16,mandra:16b}), the
effects of noise \cite{zhu:16} and chaos \cite{martin-mayor:15,hen:15a},
the effects of the underlying connectivity on the benchmarking
\cite{bunyk:14,katzgraber:14}, as well as intrinsic limitations of
transverse-field quantum annealing, such as poor performance in fair
sampling applications \cite{matsuda:09,mandra:17}.

Using the aforementioned approaches in the next sections, it is argued
that a large number of qubits is merely an engineering feat and does not
necessarily mirror a disruptive quantum optimization device. To build a
potentially disruptive device, other equally important metrics should
be considered. In what follows, the importance of these other metrics is
discussed.

\section{Appetizer --- The dream annealer}
\label{sec:hw}

Ideally, special-purpose hardware should be built with an application in
mind. This means a highly-optimized bespoke design aimed at solving a
particular problem. As such, the ``wish list'' of features a particular
device should have strongly depends on the application in mind. However,
if one may imagine a ``dream'' annealer, such device should have the
following properties:

{\em Connectivity} -- Ideally an all-to-all connectivity is desirable,
because there would be no embedding overhead \cite{comment:embedding}.
However, this is a complex task to achieve.  At the moment, connectivity
is sparse \cite{bunyk:14}. This means that multiple {\em physical}
qubits have to be used to generate {\em logical} variables and/or
couplers when the problem of interest has an underlying graph that
differs substantially from the annealer's hardware graph. As an example,
a sparse system with a fixed connectivity like in the D-Wave 2000Q
device typically requires $\sim N^2$ physical qubits to generate
approximately $N$ fully-connected logical variables
\cite{venturelli:15a}. Recent experiments on circuit fault diagnosis
problems \cite{perdomo:17y} suggest that a high connectivity might be
key in improving the performance of quantum annealing machines.

{\em Coupler order} -- At the moment, quantum annealers only permit up
to two-body qubit interactions. This means that only linear biases that
couple directly to individual qubits or qubit-qubit interactions are
possible.  However, what if the problem of interest has higher-order
qubit-qubit interactions? As an example take 3-SAT
\cite{jia:04,marques:08,biere:09,douglass:15} where each clause has
three variables. A Hamiltonian representation of the optimization
problem is ideally done with three-way couplers between the Boolean
variables. On hardware with two-way couplers, each clause would have to
be decomposed, thus requiring more physical variables.  A native
three-way qubit term in a quantum annealer would require far less
physical qubits. Because most problems of interest can be represented as
combinations of two-way, three-way and four-way couplers, the dream
annealer should have these higher-order couplers.  However, recent
experiments on circuit fault diagnosis \cite{perdomo:17y} suggest that
higher-order couplers might only show an advantage if paired with more
complex driving Hamiltonians.

{\em Better control over noise} -- Quantum annealers currently are
analog devices. Theoretically one can show that \cite{zhu:16} to first
approximation a doubling of the number of variables should be paired
with a reduction of the analog coupler noise by a factor of
approximately four. This means that for ever-growing devices, a better
control over the hardware and noise sources is needed to be able to
encode problems with high enough precision. In fact, many problems of
interest, such as the knapsack problem \cite{lucas:14}, require
precision that is currently unattainable in analog quantum annealers.

{\em Better driving Hamiltonians} -- Although this article focuses on
{\em current} hardware with transverse-field driving Hamiltonians, it
would be desirable to have higher-order non-stoquastic drivers to induce
more transitions between states and therefore potentially better
overcome barriers in the energy landscape. Recent experimental studies
have shown (see Ref.~\cite{mandra:17} and references therein) that
transverse-field quantum annealing is a biased sampling approach
\cite{matsuda:09}. The inclusion of more complex drivers might not
completely solve the sampling problem, but hopefully improve the overall
fair sampling of states.

{\em Error correction} -- Ideally, quantum error correction should be
part of any new quantum annealing machine design, especially because of
the inherent analog noise, as well as promising performance improvements
observed in previous studies \cite{pudenz:13,pudenz:15}. At the moment
error-correction schemes need to be embedded in quantum annealing
hardware, thus drastically reducing the number of available variables.
Furthermore, typically the native connectivity of the hardware is
reduced, thus making the embedding of applications onto the
error-corrected system harder. Given these limiting factors, the design
of novel quantum annealing hardware should have a built-in error
correction component via, e.g., ancilla qubits.

The fabrication of the aforementioned dream annealer with the required
specs is unlikely in the foreseeable future. While building a device
around a particular application could result in a disruptive quantum
optimization device, such application (discussed below) remains to be
found. Even worse, it is unclear how to theoretically predict if a
particular optimization problem could even benefit from quantum
optimization. Given this conundrum, in an effort to not place all eggs
in one basket, the construction of quantum annealing hardware has
focused on {\em generic} graphs with simple drivers and two-body
interactions between the qubits.

What could one do with a dream annealer and would it be better than
current CMOS technology paired with state-of-the-art optimization
algorithms? That is unfortunately unclear. In fact, there is no strong
evidence to date, that quantum annealing can excel for any application
problem beyond synthetic benchmark problems \cite{mandra:17y}. And even
for the latter the observed speedup does not warrant the effort.  So how
should the quantum annealing community proceed?

First, determine if quantum annealing can truly deliver an advantage
over classical technologies. This could either happen via theoretical
studies (desired, as long as realistic hardware considerations are taken
into account) or by the development of a (small?) {\em high-quality}
experimental test bed. An example for the former is a theoretical study
by Nishimori and Takada \cite{nishimori:17} where the effects of
non-stoquastic terms in the driver Hamiltonian are studied for different
optimization problems.  While the examples in that work are not
realistic, they do highlight the importance of analyzing a problem
before implementing it in hardware. In the case of the latter, the
IARPA-funded Quantum Enhanced Optimization Program has the goal of
addressing if quantum annealing has the potential to outperform
classical approaches provided the qubits are of the best quality
currently accessible from an engineering standpoint.  Small test beds
should ideally include the following: First, better control over the
qubits than currently available on commercial D-Wave devices. The
inclusion of more complex drivers could help elucidate if going beyond
transverse fields has potential for quantum speedup and/or better
sampling. Having quantum annealers with $k$-local ($k > 2$) topologies
would help in the understanding of the effects of embedding. A similar
argument can be made for higher connectivity. Finally, precision far
beyond the currently-available $6$ bits, (e.g., $32$ bits) would open
the doors for experimental studies on problems where high precision is
key, such as the knapsack problem. This, in turn, could assist in
finding an elusive killer application.

Second, develop the necessary know-how to assess the ``{\em quantum
potential}'' of a given application. The dissection of problems to
analyze their potential for a given class of solver/algorithm remains in
its infancy. The goal is to measure different metrics that characterize
a particular problem and determine its suitability for (in this case)
quantum architectures. Simple examples are the embedding overhead or
precision requirements. However, there could be far more subtle metrics,
such as, for example, measurements that correlate with the shape of the
cost function landscape.  Fortunately, there is an emergent industry of
companies that specialize in bridging the gap between quantum hardware
and industrial customers, such as 1QB Information Technologies,
Cambridge Quantum Computing, QC Ware, and QxBranch.

Not only is it important to understand which applications work for
quantum technologies best, it is as important to also develop sound
benchmarking strategies for quantum technologies -- the subject of the
next section.

\section{Second course --- Benchmarking lessons}
\label{sec:bench}

The most straightforward way to benchmark a novel computing paradigm
\cite{mcgeoch:12} is by determining the resources (e.g., time) needed to
meet a predefined target. There are different definitions of what this
target should be (for example, a ground state or a particular energy
value), however, there is consensus that the parameters of the device
being benchmarked should be optimized. This is of importance, because
sub-optimal parameter selection for smaller problems might lead to an
apparent better scaling of the resources needed as a function of the
problem size. As such, any claims for better performance and, in this
case, {\em quantum speedup} could not be trusted. Alternatively, one
could, for example, study the {\em quality} of the solutions found with 
a {\em fixed} amount of resources. Such an approach would, implicitly,
remove the need to optimize parameters.

Probably one of the most problematic issues in assessing the near-term
impact of quantum optimization machines is the definition of ``{\em
speedup}.'' Multiple teams have scrutinized theses devices
\cite{dickson:13,pudenz:13,smith:13,boixo:13a,ronnow:14,katzgraber:14,lanting:14,santra:14,shin:14,boixo:14,albash:15,albash:15a,katzgraber:15,martin-mayor:15,pudenz:15,hen:15a,venturelli:15a,vinci:15,zhu:16,king:17},
however, to date, it remains controversial if there is any ``{\em
quantum speedup}'' or not.  Early on it was shown that random spin-glass
benchmarks \cite{ronnow:14} on the sparse native D-Wave chimera topology
might not be complex enough to observe any advantage
\cite{katzgraber:14,katzgraber:15}. Therefore, efforts have shifted to
synthetic spin-glass benchmark problems constructed using post-selection
\cite{katzgraber:15,zhu:16,marshall:16} methods, planted solutions
\cite{hen:15a,king:17} or gadgets \cite{denchev:16,albash:17}.  Although
some of the aforementioned results suggest that the commercially
available D-Wave device has some advantage for carefully-designed
synthetic problems, this advantage often was a constant speedup over
other classical approaches, i.e., a similar scaling with the size of the
input. Even worse, for a wide variety of these gadgets the logical
structure of the underlying problem is easily decoded (e.g., via a
decimation heuristic) and the remaining logical problem solved in
polynomial time \cite{mandra:16b,mandra:17a} with exact methods.
Therefore, any speedup claims are tentative, at best.

{\em Definitions of quantum speedup} -- The first careful definition of
quantum speedup was done in Ref.~\cite{ronnow:14a}. In particular, the
authors of Ref.~\cite{ronnow:14a} differentiate between the following
categories: {\em Provable quantum speedup}, {\em strong quantum speedup}
(comparison to the best classical algorithm, regardless if the algorithm
exists or not), {\em potential quantum speedup} (comparison to the best
known classical algorithms), as well as {\em limited quantum speedup}
(in this case, comparison against simulated quantum annealing via
quantum Monte Carlo).  In an effort to add more granularity to
benchmarks, Ref.~\cite{mandra:16b} introduced the notion of {\em limited
sequential quantum speedup}, where comparisons are restricted to the
best known {\em sequential} algorithms. Within this class, quantum
annealers, as well as classically-simulated quantum annealers currently
outperform all other known sequential algorithms for different benchmark
problems.

{\em Slope vs offset} -- What would constitute a disruptive advantage of
quantum annealers over classical optimization techniques? Purists might
argue a change in the scaling (slope). This means that when analyzing
the resource requirements (e.g., time to solve a problem) as a function
of the number of variables, the growth of the requirements should be
less pronounced than for classical hardware. However, what if the
scaling is not better, but there is a constant offset that is several
orders of magnitude? Likely, a constant offset of $10^4$ would not be
disruptive, because, using a parallel implementation of a classical
algorithm on $10^4$ compute cores is readily available, i.e., no
disruptive advantage. However, what if this constant offset is
$10^{12}$? In that case, not only would we be out of luck with classical
hardware, there would even be enough wiggle room to break up large
problems that require more than $1000$ qubits and solve them on the
quantum hardware \cite{zintchenko:15}. Still, to assess quantum speedup
in the traditional sense, an improvement in the scaling is expected
\cite{comment:power,comment:optimal}.

{\em Synthetic vs application speedup} -- The notion of speedup also
strongly depends on the benchmark problem used. Although there have been
indications of benchmark problems where the D-Wave 2000Q device
\cite{mandra:17y} outperforms all known classical algorithms (without a
visible scaling advantage), these are synthetic problems designed to
``break'' all known classical algorithms. A real-world application where
quantum annealing outperforms all known classical algorithms remains to
be found and would constitute a strong endorsement for quantum annealing
to be a potentially disruptive technology.

Summarizing, it is argued that any claims for speedup should include
{\em all} algorithms that are known to be amongst the best (see
Refs.~\cite{hartmann:01,juenger:01,hartmann:04,mandra:16b,mandra:17a,mandra:17y}
for an overview), as well as benchmarks from actual application
problems. However, because application problems with a potential for a
quantum advantage remain to be found, the development of synthetic
spin-glass
benchmarks will play a predominant role in the field.  These should
ideally have planted solutions, tunable typical computational
complexity, impossible to deconstruct by tailored algorithms, and known to
be hard for all currently known classical solvers \cite{mandra:17y}.
The interplay between application-based and synthetic benchmarks,
careful design of benchmarking strategies, clear definitions of
speedup, as well as diligent use of statistics are key in assessing any
potential future quantum speedup over classical technologies.

\section{Entremet --- Other benchmarking ideas}
\label{sec:other}

Because quantum annealers are designed to solve hard optimization
problems fast, the benchmarking focus has been almost exclusively on
speed. However, there are applications---for example molecular
similarity in quantum chemistry \cite{hernandez:17}, probabilistic
membership filters \cite{weaver:14}, or machine learning---where a
spectrum of uncorrelated solutions are more important than the actual
optimum of the cost function. It was recently shown that vanilla quantum
annealing with a transverse-field driver is a rather biased sampler
\cite{mandra:17}.  If quantum annealing machines are expected to
efficiently tackle these problems in the near future, efforts should
shift to mitigate this currently existent exponential bias towards a
subset of states.  Therefore, in addition to assessing the speed of
quantum annealing machines, an independent metric should be the fair
sampling \cite{mandra:17} capability of a particular device. Assuming
near-uncorrelated solutions, a random unbiased sampler should find each
solution of a degenerate problem with approximately the same probability
(up to statistical fluctuations) \cite{moreno:03}.  Combined with sound
quantum speedup benchmarks, this would represent a benchmarking gold
standard for new hardware that uses different driving Hamiltonians,
post-processing schemes, as well as applications that require different
uncorrelated solutions.

\section{Main course --- Applications}
\label{sec:applications}

As outlined in the introduction, multiple optimization problems across
disciplines can be mapped onto spin-glass-like binary problems
\cite{lucas:14}, up to a potential embedding overhead. If the community
develops a notion of which types of problems have potential for a
quantum advantage, tailored quantum optimization machines could be
designed to tackle these. However, despite all efforts, a ``killer''
application or problem domain where quantum optimization excels has yet
to be found.

{\em What has been done?} -- NASA's Quantum Artificial Intelligence Lab
has been a pioneer in the study of applications on current quantum
annealing hardware. These applications range from spin glasses
\cite{venturelli:15a}, to lattice protein models \cite{perdomo:12}, fault
diagnosis in graph-based systems \cite{perdomo:15b} (e.g., circuits
fault diagnosis \cite{perdomo:17y}), operational planning
\cite{rieffel:15}, job-shop scheduling \cite{venturelli:15b}, and
quantum-assisted machine learning \cite{benedetti:16}, to name a few. In
addition, different corporations such as 1QB Information Technologies
(1QBit) have emerged, that aim at bringing novel computing techniques --
such as quantum computing -- to different enterprises. For example,
1QBit has studied different customer-driven application problems such as
molecular similarity in chemistry \cite{hernandez:17} or the optimal
trading trajectory problem \cite{rosenberg:16}. These efforts have the
largest potential in identifying problems that are well-suited for
quantum annealing technologies, as well as for future digital
programmable quantum computing systems.

{\em Where vanilla quantum annealing likely won't work} -- Because
transverse-field quantum annealing is a biased sampler
\cite{matsuda:09,mandra:17}, it is unlikely that without fundamental
changes to the hardware and driving Hamiltonians quantum annealing will
have an impact in applications that require an extensive sampling of a
solution space with degenerate solutions. This means that current
transverse-field implementations are unlikely to have any transformative
impact in applications such as machine learning, inference, image
recognition, probabilistic membership filters using SAT
\cite{weaver:14}, or the optimization of the geometry of molecules in
chemistry applications. However, future generations of these devices
might include either better driver Hamiltonians or corrective
post-processing schemes that might alleviate this problem. A recent
example for post-processing of data was introduced in
Ref.~\cite{ochoa:18x} and alleviates the biased sampling found in the
D-Wave device.

{\em Where the vanilla approach might work} -- Recently, D-Wave Systems
Inc.~demonstrated that their quantum annealing hardware can be used as a
{\em physical simulator} \cite{katzgraber:14,harris:17x}. In that study,
the machine's parameters were tuned to simulate a physical
three-dimensional Edwards-Anderson Ising spin glass \cite{edwards:75}.
Current hardware with approximately $2000$ qubits allows for the
embedding of a three-dimensional quantum spin glass of $512$ variables.
The simulation of a quantum spin glass on traditional CMOS technology at
low temperatures and with large transverse fields is notoriously
difficult.  By doubling the number of variables in the hardware the
quantum annealer would be able to solve quantum spin-glass problems in a
regime hard to probe for classical computers.  Similarly, the machine
could be used to study other phases of matter, such as frustrated
pyrochlore magnets \cite{berg:03}, simply by treating the Boolean
variables as physical object.  Although at first sight the application
scope seems limited, not only are there multiple problems across
condensed matter and statistical physics that should be revisited on an
analog quantum simulator, but this would be the fulfillment of Richard
Feynman's dream of a quantum simulator \cite{feynman:85}.

{\em So how should the community proceed?} First, a clear establishment
of what properties of a problem and/or application make it well-suited
for quantum optimization should be established. Tackle the quantum
speedup problem from the bottom up by {\em first} understanding what
makes a problem well-suited for quantum annealing and {\em then}
building a special-purpose machine to solve it.  In the medium to long
term this will carve out the application scope of quantum annealing, as
well as drive research and development in the right directions.
Finally, although one could argue that simulated quantum annealing on
CMOS technology might scale better than an analog quantum device, the
offset between heuristics on CMOS and quantum hardware implementations
remains huge. For example, in a recent study \cite{denchev:16} the
difference was approximately a factor of $10^7$.  Taking into account
power consumption, this represents a sizable advantage for quantum
annealing machines.

\section{Cheese plate --- Benefits for classical?}
\label{sec:classical}

It is unlikely that classical CMOS technology will be fully replaced by
quantum computing machines. For the next few decades, standard CMOS
technology will still power most of humanity's computing needs from
smartphones to servers while quantum computers will likely serve as large-scale
co-processors in the cloud used for problems where classical would
either require huge resources or quantum is optimally suited. In the
near term, however, developments in quantum optimization have led to an
``{\em arms race}'' with classical technologies. Not only have there
been many algorithmic advances, there has been a clear shift to custom
silicon hardware.

Algorithmically, heuristics from the study of spin glasses have been
adopted in the field of quantum optimization and, subsequently, by
corporations interested in quantum and quantum-inspired optimization.
State-of-the-art simulation approaches commonly used in the study of
spin glasses are being adopted by industry. For example, replica
exchange methods such as parallel tempering Monte Carlo
\cite{hukushima:96,katzgraber:06a}, particle swarm methods such as
population annealing Monte Carlo \cite{hukushima:03,wang:15,wang:15e},
and cluster algorithms such as isoenergetic cluster moves \cite{zhu:15b}
have been adopted by companies specialized in the optimization of binary
problems \cite{karimi:17}.  In parallel, the emulation of quantum
approaches on traditional hardware has led to the development of
highly-efficient heuristics, such as simulated quantum annealing, that
are becoming more commonplace both in academic, as well as industrial
settings.

On the hardware side and driven by artificial intelligence applications,
corporations are increasingly using field-programmable gate arrays
(FPGAs) or custom hardware, such as Fujitsu's Digital Annealer
\cite{matsubara:18} or Google Inc.'s Tensor Processing Unit,
respectively. Furthermore, Microsoft has invested heavily in the
development of highly-interconnected FPGAs via Project Catapult, a
configurable cloud architecture. Such custom hardware cloud processing
schemes set the stage for the eventual inclusion of quantum
co-processors in the cloud.  In the near-term, a hybrid
classical-quantum cloud framework could be built with specific
applications in mind. Eventually, once researchers are able to
``predict'' if a particular application is well suited for quantum
approaches, more generic hybrid frameworks could be deployed. As
emphasised in Sec.~\ref{sec:applications}, gaining a deeper
understanding of what makes a problem amenable for quantum
(optimization) approaches should be in the center of current quantum
computing research and hardware developments.

\section{Dessert \& Digestif --- Concluding remarks}
\label{sec:conclusion}

The development of synthetic benchmarking strategies based on Boolean
frustrated systems will play a pivotal role in the development of
near-term quantum optimization machines. Furthermore, tools from the
study of disordered and frustrated systems will play an increasing role
in assessing the quantum-readiness of particular quantum computing
applications. For the near term, quantum optimization using
transverse-field quantum annealing on quadratic Hamiltonians will remain
the standard. With all the insights gained by multiple researchers
across fields, technology will be advanced further and---hopefully
soon---to a point where quantum optimization has a clear application
scope and outperforms classical computing technologies. If no true
computational advantage can be found, either via a strong scaling
advantage for a given application or a constant offset with a
considerably smaller energy footprint than classical computing
technologies, quantum annealing with a transverse field will likely fade
away and be superseded by digital devices.  Finding applications with
even the weakest quantum speedup could give the quantum optimization
field a massive boost that, in turn, will leverage other quantum
computing developments, e.g., digital (where researchers will learn
valuable lessons from analog developments). And these quantum computing
developments will henceforth bolster the development of new classical
computing technologies. It is this arms race that will herald a
computing revolution.

\subsection*{Acknowledgments}

H.G.K.~would like to thank his current and former team members,
A.~Barzegar,
J.~Chancellor,
C.~Fang,
D.~C.~Jacob,
O.~Melchert,
H.~Munoz-Bauza,
A.~J.~Ochoa,
C.~Pattison,
W.~Wang,
and,
Z.~Zhu,
as well as multiple collaborators,
S.~Mandr{\`a}, A.~Perdomo-Ortiz, E.~Rieffel (NASA QuAIL), 
M.~Aramon, L.~Downs, A.~Fursman, M.~Hernandez, H.~Karimi, G.~Rosenberg (1QBit),
F.~Hamze (D-Wave Systems Inc.), 
B.~Jacobs, K.~F.~Roenigk (IARPA),
W.~Oliver, A.~J.~Kerman (MIT \& MIT Lincoln Laboratory),
J.~Machta (University of Massachusetts Amherst),
M.~A.~Novotny (Mississippi State University),
C.~K.~Thomas (Google Inc.),
and,
M.~Troyer (Microsoft) 
for multiple discussions and fruitful collaborations.
H.G.K.~acknowledges support from the National Science Foundation (Grant
No.~DMR-1151387) and would like to thank the culinary world for
inspiration.  This work is supported in part by the Office of the
Director of National Intelligence (ODNI), Intelligence Advanced Research
Projects Activity (IARPA), via MIT Lincoln Laboratory Air Force Contract
No.~FA8721-05-C-0002. The views and conclusions contained herein are
those of the authors and should not be interpreted as necessarily
representing the official policies or endorsements, either expressed or
implied, of ODNI, IARPA, or the U.S.~Government.  The U.S.~Government is
authorized to reproduce and distribute reprints for Governmental purpose
notwithstanding any copyright annotation thereon.

\vspace*{1em}
\hrule
\vspace*{1em}

\bibliographystyle{iopart-num.bst}
\bibliography{comments,refs}

\begin{thebibliography}{10}
\bibitem{comment:embedding}
Withing this context, an embedding of a graph $G$ into the graph $H$ of quantum
  hardware is a subgraph of $H$, such that $G$ can be obtained from $H$ by
  contracting edges.
\bibitem{comment:optimal}
A careful scaling analysis requires the precise optimization of all parameters
  in the optimization approach. Clearly, this poses a problem, because the
  optimization has to be run multiple times with different parameter settings
  to obtain the optimal set. Needless to say, this parameter optimization
  procedure should be included in the benchmarks, however, this is usually not
  the case. An alternative would be to use other time-independent metrics, such
  as residual energies or relaxed benchmarking approaches \cite{katzgraber:15}
  that take the analog nature of quantum annealing machines into account.
\end{thebibliography}

\end{document}